\documentclass[11pt]{article}
\usepackage{amsmath, amssymb}
\usepackage{times}
\usepackage{graphicx}
\usepackage{xcolor}
\usepackage{multirow}
\usepackage[authoryear]{natbib}
\usepackage{rotating} 
\usepackage{soul}
\usepackage{comment}
\usepackage{bbm}  
\usepackage{latexsym} 
\usepackage{scrextend}
\usepackage[margin=1.2in]{geometry}

\usepackage{hyperref}
\newcommand{\footremember}[2]{%
    \footnote{#2}
    \newcounter{#1}
    \setcounter{#1}{\value{footnote}}%
}
 
\begin{document}

\title{Computationally efficient model selection for joint spikes and waveforms decoding}

\author{%
  Francesca Matano\footremember{alley}{Department of Statistics, Carnegie Mellon University, Pittsburgh, PA 15213}\\
  \texttt{fmatano@andrew.cmu.edu}\\
  \and Val\'erie Ventura\footremember{trailer}{Department of Statistics, Carnegie Mellon University, Pittsburgh, PA 15213, and the Center for the Neural Basis of Cognition, Pittsburgh, PA 15213}\\
  \texttt{vventura@stat.cmu.edu}
  }


\date{}
\maketitle

\noindent



\begin{abstract}
A recent paradigm for decoding behavioral variables or stimuli from neuron ensembles 
relies on joint models for electrode spike trains and their waveforms,
which, in principle, is more efficient than decoding from electrode spike trains alone or from sorted neuron spike trains.
In this paper, we decode the velocity of arm reaches of a rhesus macaque monkey to show 
that including waveform features indiscriminately in a joint decoding model
can contribute more noise and bias than useful information about the kinematics, and thus degrade decoding performance.
We also show that selecting which waveform features should enter the model to lower the prediction risk
can boost decoding performance substantially.
For the data analyzed here, a stepwise search for a low risk electrode spikes and waveforms joint model
yielded a low risk Bayesian model that is 30\% more efficient than the corresponding risk minimized 
Bayesian model based on electrode spike trains alone. 
The joint model was also comparably efficient to decoding from a risk minimized model based only on 
sorted neuron spike trains and hash, confirming previous results that one can do away with the 
problematic spike sorting step in decoding applications.
We were able to search for low risk joint models through a large model space thanks to a
short cut formula, which accelerates large matrix inversions in stepwise searches
for models based on Gaussian linear observation equations.
\end{abstract}



\clearpage

\section{Introduction}

A wide range of models have been used successfully to decode behavioral variables or other stimuli encoded by
neuron ensembles, from electrode signals recorded by microelectrode arrays.
A popular choice is to avoid spike sorting and treat electrodes as single putative neurons, which is fast and thus desirable for clinical use \citep{Fraser2009}, but ignores the kinematic information provided by individual neurons.
But spike sorting is difficult, imperfect, and time-consuming, and when it is performed without considering the covariates that modulate neurons spiking,
estimates of neurons' tuning functions are biased in a way that degrades decoding \citep{Ventura2009b}.
Noting that waveform based spike sorting provides information about rate estimation and vice versa,  
\cite{Ventura2009a, Ventura2009b} suggested that 
one should consider the two relationships simultaneously rather 
than sequentially and indeed observed better decoding performance in simulated data 
when using a joint model for electrode spike trains, waveforms, and kinematics to first soft 
sort the electrode spike trains, and use these sorted data to decode.
\cite{Todorova2014a} decoded kinematics directly from the same joint model, without the
soft sorting intermediate step, and \cite{Kloosterman2013} proposed a fully nonparametric implementation 
of the same approach.
Both these implementations are computationally intensive, and thus a detriment to clinical applications of decoding. \cite{Ventura2015} thus proposed a fully parametric implementation based on linear Gaussian observation equations, which yields closed-form optimal linear predictions (OLE, \cite{Salinas}) and Bayesian predictions \citep{Brown1998}.
With computational efficiency in mind, we adopt this implementation in this paper.

\cite{Ventura2015} further relied on the information processing inequality to argue that the 
electrode spike trains and their spike waveforms together contain more information about the kinematics 
than the electrode spike trains alone or the sorted neurons spike trains, so that decoding jointly from 
the electrode spike trains and waveforms is most efficient, in principle.
%
%
%
\cite{Kloosterman2013} reduced the waveforms recorded by tetrodes to the
four-dimensional vector of their amplitudes on each electrode and reported a 14\% mean improvement for
OLE decoding compared to decoding from sorted units with hash, to reconstruct the 1D location of rats from hippocampal place cells. 
%
\cite{deng2015} used the same implementation in a Bayesian
framework to decode the 1D location of rats from tetrodes implanted in the hippocampus, also reducing the 
waveforms their four amplitudes, and observed a performance similar to decoding from manually sorted 
units with hash.
\cite{Todorova2014a} decoded the 3D arm velocity of two rhesus macaques from PvM microarray data, reducing the waveforms to their amplitudes.
The joint OLE model provided a 19\% improvement over the corresponding model based on manually
sorted neurons with hash, for one monkey. The two models performed similarly for 
the other monkey and the two corresponding Bayesian models performed similarly for both 
monkeys. Compared to decoding from unsorted electrodes, the joint model was 13 to 28\% more efficient for
OLE decoding, and 2 to 7\% for Bayesian decoding.
%
How well electrodes are sorted and how the joint model is built matters.
\cite{Ventura2015} used a different, selected joint model (see next paragraph) to decode the same data, which outperformed
decoding from the manually sorted neurons with hash, within OLE, Bayesian, and forward filter decoding paradigms,
and that was comparable to decoding from hash and neurons sorted using the expert sorter of \cite{Carlson2014}.
Decoding from unsorted electrodes was least efficient within all paradigms.

The applications above reduced the waveforms to their amplitudes and included 
the amplitudes of all the electrodes in the decoding model.
But waveforms do not all provide the same information.
\cite{Ventura2015} argued that waveforms from electrodes that record many neurons (e.g. tetrodes) 
that have different tuning curves contain substantial information that is unlikely to be fully captured by a single waveform 
feature (indeed \cite{Kloosterman2013} observed an efficiency  bump of 26\% from using two of the tetrode 
amplitudes instead of one, and a further 14\% bump from using four instead of two), and that conversely, 
waveforms provide no information in addition to the information in spike trains, for electrodes that record 
only one neuron or several neurons that are similarly modulated by the kinematics.
Yet, including waveforms in the decoding model injects noise in the predicted kinematics
since their models are estimated from training data, and bias when parametric models are used
\citep{Todorova2017}.
Hence, there is a tradeoff between the amount of information provided by the waveforms 
and the amount of additional variability and bias they induce.
Model selection methods are needed to balance these elements,
with objective to minimize the prediction risk.
\cite{Ventura2015} reduced the waveforms to the four features shown in Fig.~\ref{fig:wf_feat} and decoded from two joint models:
the first included the first three moments of the first feature (the amplitude) of all electrodes, and the second,
selected moments amongst the first five moments and first cross-moments of the four features.
Decoding from the second model was as efficient as decoding from expertly sorted neurons
with hash within OLE, Bayesian, and forward filter decoding paradigms. The first joint model was about 
10\% less efficient in median across a training set of arm reaches.
%

Model selection is straightforward for forward filters that predict kinematics as functions of neural covariates, using 
stepwise regression or penalized approaches like LASSO or ridge regression, and can improve decoding accuracy very substantially. 
Model selection for OLE and Bayesian models can also rip substantial rewards, but a major hurdle is the computational burden of recomputing the prediction risk for each candidate model, which prevents 
stepwise, let alone exhaustive searches \citep{Todorova2017}.
In their stepwise search, \cite{Ventura2015} and \cite{Todorova2017} selected waveform features using an added variable test (AVT) 
that prevented highly correlated observation equations from entering the joint model, 
which had the advantage of speed but was not particularly effective to reduce the prediction
risk.

In this paper we develop a short cut formula to recalculate OLE and Bayesian predictions from 
decoding models based on Gaussian linear observation equations, 
which allows computationally efficient stepwise searches to identify low risk models,
and apply the method to decode the 3D velocity of the arm reaches of a monkey.
We show that including in the joint decoding model
waveform features from all the electrodes can contribute more noise and bias than useful information about the kinematics, and that
searching for low risk models through a large model space can boost decoding performance substantially.

\section{Methods}

\paragraph{Spikes and waveforms joint encoding model}

\begin{figure}
\centerline{ \includegraphics[scale=0.3]{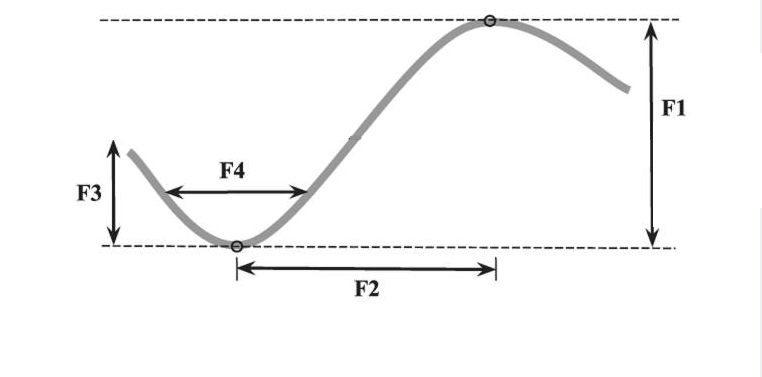}}
\vspace{-.2in}
\caption{\sl Waveform features whose moments we consider: F1: amplitude; F2: time elapsed from peak to trough voltage; F3: size of trough; F4: width of trough at half minimum.}
\label{fig:wf_feat}
\end{figure}

Assume that we record the signals of $n$ electrodes to decode the $p$-dimensional
kinematics $\mathbf{k}_{t}=(k_{1t}, \ldots, k_{pt})$ at time $t$; $p=3$ in the results section.
For each electrode, we record the time of occurences and associated waveforms of
all supra-threshold events, be they spikes or noise; we refer to them as spikes for brevity.
We let $\mathbf{s}_t = \left(s_{1t}, \ldots, s_{nt}\right)^T$ be the $n \times 1$ vector of the $n$ electrodes' spike counts
in the bin of size $\delta$ centered at $t$,
and 
$\overline{\mathbf{w}_{lt}^m} =  \left(  \overline{w^m_{1lt}}, \ldots, \overline{w^m_{nlt}}   \right)^T$ be the $n \times 1$ vector of the $m$th moments of the $l$th waveform feature 
for each electrode, where the $j$th coordinate
\begin{equation}
\overline{w^m_{jlt}} = \delta^{-1} \sum_{i=1}^{s_{jt}} w_{jlit}^m
\label{moment}
\end{equation}
is the sum of the $s_{jt}$ observed occurrences on electrode $j$, in the bin of size $\delta$ centered at $t$, 
of the $l$th waveform features exponentiated to the $m$.
\cite{Ventura2015} argued that the joint distribution of waveforms can be fully represented by the set of moments of all orders; to work with a finite set, they reduced the waveforms to the $L=4$ features shown in Fig \ref{fig:wf_feat} 
and used their moments up to order $M=5$ and cross-moments of order two. 
These features can be collected in real time.
The methods discussed here apply similarly to other features or summaries, e.g. PCs,
or to each coordinate of unreduced waveforms, but this requires that waveforms be aligned, 
so the electrode signals cannot be discarded in real time.

The formulation of the joint  spikes and waveforms decoding framework in \cite{Ventura2015} is as follows.
The lagged spike counts are assumed to be Gaussian variables with firing rates linear in the kinematics:
\begin{equation}
\label{ObsMod}
\mathbf{s}_{t-\tau} = \mathbf{\alpha} + \mathbf{C} \ \mathbf{k}_{t} + \mathbf{\eta}_t ,
\end{equation}
where $\alpha$ is a $n \times 1$ vector and $\mathbf{C}$ a $n \times p$ matrix of regression coefficients, and
$\mathbf{\eta}_t$ is a $n \times 1$ Gaussian noise 
vector with covariance matrix $\mathbf{Z}$, where $\alpha$, $\mathbf{C}$, and $\mathbf{Z}$ are fitted to training 
data by maximum likelihood (ML). 
Under the linearity assumption in eq.\ref{ObsMod}, the waveform moments are also Gaussian and linear \citep{Ventura2015}, 
so we can write:
\begin{equation}
\label{ObsModW}
\overline{\mathbf{w}_{l (t-\tau)}^m} = \mathbf{\gamma}_{lm} + \mathbf{G}_{lm} \ \mathbf{k}_{t} + \mathbf{\delta}_{klt} ,
\end{equation}
where $\gamma_{lm}$ is a $n \times 1$ vector and $\mathbf{G}_{lm}$ a $n \times p$ matrix of regression coefficients
fitted to training data by maximum likelihood (ML).
The covariance matrix of the $n \times 1$ Gaussian noise $\mathbf{\delta}_{lmt}$ depends on the observed spike counts since 
eq.~\ref{moment} depends on them.
Therefore, for each electrode $j$, feature $l$, and moment $m$, 
given that we observe $s_{jt}$ spikes for electrode $j$ in bin $t$, the variance of the $m$th moment of the $l$th waveform feature 
in bin $t$ is 
$$
\mathbb{V}(\overline{w^m_{jlt}}) = \delta^{-2} \, s_{jt} \,  \mathbb{V}( w^m_{jlit} ),
$$
 where $\mathbb{V}( w^m_{jlit} )$ is
the variance of the $l$th waveform feature exponentiated to the $m$, which we estimate with the sample 
variance of all its occurrences in the training dataset (this makes the reasonable assumption that the variance 
is constant and does not depend, for example, on the number of spikes in bins).
Similarly, for each $j, q, l, h, m$ and $n$, the covariance between waveform moments is
$$
\mathbb{C}ov(\overline{w^m_{jlt}}, \overline{w^n_{qht}}) = \delta^{-2} s_{jt} s_{qt} \mathbb{C}ov(w^m_{jlit}, w^n_{qhrt})
$$
where the covariance between the same ($n = m$ and $l=h$) or different ($n \neq m$ or $l\neq h$) features recorded on the same ($j=q$) or different ($j \neq q$) electrodes is estimated as the sample covariance of the corresponding observations in the training dataset.
Finally, the covariance between spike counts $s_{jt}$ and waveform feature $\overline{w^n_{qht}}$ is always zero since, given $s_{jt}$, $s_{jt}$ is a constant.

\paragraph{OLE and Bayesian decoding}

Let $\mathbf{c}_{t}$ be an $N \times 1$ vector composed of a subset of, or all the spike counts $\mathbf{s}_{t-\tau}$ and waveforms moments $\overline{\mathbf{w}^m_{l(t-\tau)}}$ in eqs.~\ref{ObsMod} and~\ref{ObsModW}.
The joint decoding model of \cite{Ventura2015} consists of the corresponding $N$ observation equations:
\begin{equation}
\label{ObsModFull}
\mathbf{c}_t = \mathbf{\beta} + \mathbf{B} \ \mathbf{k}_{t} + \mathbf{\zeta}_t ,
\end{equation}
where the model parameters $\beta$, $\mathbf{B}$, and covariance matrix $\mathbf{U}$ of $\mathbf{\zeta}_t$ are 
composed of the corresponding model components in Eqs.\ref{ObsMod} and~\ref{ObsModW}.
Given eq.~\ref{ObsModFull} and new neural data $\mathbf{c}_t$, the optimal linear 
estimator (OLE) is obtained in closed form as the MLE of $ \mathbf{k}_{t}$ in eq.\ref{ObsMod}:
\begin{equation}
\mathbf{\tilde k}_{t}^{OLE} =\mathbf{M}^{OLE} ( \mathbf{c}_{t}  - \mathbf{ \beta}),
\label{eq:OLEpred}
\end{equation}
where 
\begin{equation}
\mathbf{M}^{OLE}  = ( \mathbf{ B}^{\prime}  \mathbf{ U} ^{-1} \mathbf{ B})^{-1}   \mathbf{ B}^{\prime}  \mathbf{U} ^{-1}.
\label{eq:MOLE}
\end{equation}
Bayesian decoding supplements eq.\ref{ObsMod} with a kinematic evolution equation, here the autoregressive process of order one:
\begin{equation}
\label{StateEq}
\mathbf{k}_t = \mathbf{A} \ \mathbf{k}_{t-1} + \mathbf{\epsilon}_t ,
\end{equation}
where $\mathbf{A}$ is a $p \times p$ matrix of coefficients and $\mathbf{\epsilon}_t$ are Gaussian perturbations
with covariance matrix $\mathbf{W}$, with $\mathbf{A}$ and $\mathbf{W}$ fitted to training data.
The Bayesian kinematic prediction is given by the Kalman recursive equations:
\begin{equation}
\mathbf{\tilde k}_{t}^{KF} =  \mathbf{M}^{KF}_t  ( \mathbf{c}_{t} -\beta ) + (  \mathbf{I} - \mathbf{M}^{KF}_t   \mathbf{ B}) \mathbf{ A} \mathbf{\tilde k}_{t-1}^{KF} 
\label{eq:KFpred}
\end{equation}
where 
\begin{equation}
\mathbf{M}^{KF}_t = \mathbf{ \Sigma}_{t \mid t-1} \mathbf{B}^{\prime} ( \mathbf{B \Sigma}_{t \mid t-1} \mathbf{B}^{\prime} + \mathbf{U)}^{-1},
\label{eq:MKF}
\end{equation}
$ \mathbf{ \Sigma}_{t \mid t-1} $ satisfies the Riccati recursion \citep{Kalman1960,Brown1998},
%
with initial condition $ \mathbf \Sigma_{0 \mid -1}  =  \mathbf \Sigma_0$, the covariance matrix of the 
initial velocity $\mathbf{v}_{0}$. We initiate the KF decoder 
at the true velocity and set $\mathbf \Sigma_0 = 0$.

Our goal in this paper is to identify minimum or low risk OLE and Bayesian decoding models, which consists of 
identifying which observation equations to include in eq.~\ref{ObsModFull} \citep{Todorova2017}. 
To avoid overfitting, we score each model by calculating a cross-validated estimate of the prediction 
risk in a training set, with prediction risk taken to be the mean squared error (MSE), 
and we report the prediction risks evaluated on a separate testing set in our figures and comments.
Details about training and testing sets, and cross-validation are in the results section.

Next, we augment the model space to hopefully include additional low risk decoding models,
and we discuss how to implement an efficient stepwise search of the model space.

\paragraph{Model space}

At present, the model space is composed of all subsets of the observation equations in Eqs.\ref{ObsMod} and~\ref{ObsModW}.
The lag $\tau$ between neural activity and kinematics is often taken to be the same across all electrodes
or all neurons, but \cite{Wu2006} and \cite{Todorova2017} show that using different lags can improve decoding performance. We therefore consider that option, allowing temporal lags between neural data and kinematics to be different across electrodes, from $\tau=0$ to $12$ time bins, where $\tau=12$ corresponds to 192 ms with the 16 ms wide bins we use in the results section. We proceed similarly with all waveform moments.

Next, the assumed parametric models in Eqs.\ref{ObsMod} and~\ref{ObsModW} are biased 
since they cannot match exactly the true models, and this in turn biases kinematic predictions.
Following \citep{Todorova2017}, we augment the model space with alternative conditional expectation (ACE, \cite{Breiman1985}) response transformation models:
\begin{equation}
\label{ObsMod2}
g(\mathbf{s}_{t-i}) = \mathbf{\alpha}_{gi} + \mathbf{C}_{gi} \ \mathbf{k}_{t} + \mathbf{\eta}_{git} , \quad i=0, 1, \ldots, 12,
\end{equation}
where $g$ is a nonparametric function that maximizes the correlation between the left and right hand sides of eq.\ref{ObsMod2}.
We fit a specific ACE function $g$ for the spike counts of each electrode at each lag $i$, using the R function {\tt ace} in package {\tt acepack}.
We proceed similarly for the waveform moment equations in eq.~\ref{ObsModW}.
The ACE transformed equations are more flexible than their linear counterparts in eqs.~\ref{ObsMod} and 
eq.~\ref{ObsModW} so they are more flexible, yet they remain linear in the kinematics so they still 
yield closed form OLE and Bayesian predictions.
They are less biased, but the added flexibility comes at the 
cost of increased variance, so they may not always provide a better bias variance tradeoff. We therefore keep 
both linear and ACE equations in the model space and will let the model selection procedure determine 
which are most useful.
We further include square root response transformation equations, that is eq.\ref{ObsMod2} with $g(\cdot)= \sqrt{\cdot}$, and similarly for waveforms, because square root spike counts
can be better linearly related to the kinematics than raw counts \citep{Moran1999,Wu2006}
and thus have lower bias than linear equations, and lower variance than nonparametric equations.

To summarize, our model space includes the decoding models composed of 
subsets of the observation equations. Each electrode contributes 
one linear equation for the raw spike counts for each of 13 lags, 2x13 more equations for the square root and ACE transformed spike counts, and the corresponding 13x3 linear equations for each of the 4x2 waveform moments (the 2 first moments of 4 waveform features), which makes a total of 351 available observation equations per electrode.
However, we constrain all decoding models to include, per electrode, at most one transformed or untransformed spike count equation, one equation for each of the first and second moments of the first waveform feature, 
and similarly for the other three features, so that an electrode can contribute at most nine observation equations 
to a decoding model. Otherwise, stepwise OLE model
searches would keep adding equations to the model to lower the risk by smoothing kinemetic predictions,
much like Bayesian predictions do \citep{Todorova2017}.
This phenomenon is of no particular interest since one can obtain smooth trajectories from a smaller model 
by using either a Bayesian decoder or an OLE decoder with smoothed neural data.


\paragraph{Computationally efficient stepwise model searches  \label{sec:OLEmodsel}}

An exhaustive search through the model space to identify the minimum risk model is computationally prohibitive, as 
are stepwise searches to identify low risk models \citep{Todorova2017}.
Below we derive a short cut formula to recalculate the kinematics more efficiently in stepwise searches, 
which in turn makes it more efficient to evaluate the risk of candidate models. The code is available at https://github.com/fmatano/KalmanFilteR and https://github.com/fmatano/State-Space-Stepwise.

Assume that we want to add or remove $k$ observation equations to or from a current model of $N$ equations.
The derivations below are valid for all integers $k$ although we used $k=1$ in the results section.
To recompute OLE prediction in eqs.~\ref{eq:OLEpred}, we trivially add/remove the appropriate elements
to/from $\mathbf{c}_{t}$, $\mathbf{ \beta}$, and $\mathbf{ B}$.
However, inverting the appropriately augmented or reduced matrix $\mathbf U$ in eq.~\ref{eq:MOLE}
is computationally expensive when $N$ is large, with cost in $O(N^3)$.
Calculating $(\mathbf{ B}^{\prime}  \mathbf{U}^{-1} \mathbf{ B})$ once $\mathbf U^{-1}$ is 
available involves multiplying large matrices, which is comparably faster with cost in $O(N^2)$,
and inverting $(\mathbf{ B}^{\prime}  \mathbf{U}^{-1} \mathbf{ B})$ is trivial since it has dimension $p \times p$ 
where $p$, the dimension of the kinematics, is typically less than three.
Recomputing the Bayesian prediction in eq.~\ref{eq:KFpred} proceeds similarly, where the computationally
most expensive step is the inversion of the appropriately augmented matrix
$(\mathbf{B \Sigma}_{t \mid t-1} \mathbf{B}^{\prime} + \mathbf{U})$ in eq.~\ref{eq:MKF}. 
%
%
%

Let ${\mathbf V}$ be an $N \times N$ symmetric matrix; the derivations below apply similarly 
to $\mathbf{V = U}$ in eq.~\ref{eq:MOLE} for OLE predictions, and to
${\mathbf V =  (\mathbf{B \Sigma}_{t \mid t-1} \mathbf{B}^{\prime} + \mathbf{U})}$ in eq.~\ref{eq:MKF} for Bayesian predictions.
Let $\mathbf{V}_{\pm k}$ denote the appropriately reduced or augmented matrix ${\mathbf V}$.
The short-cut formula derived below cuts down on computation cost by updating  ${\mathbf V}^{-1}$ to obtain 
$(\mathbf{V}_{\pm k})^{-1}$ rather than inverting $\mathbf{V}_{\pm k}$ directly; it has computation 
cost in $O(N^2)$ rather than $O(N^3)$.

Assume first that we want to remove $k$ out of the $N$ observation equations. We re-order the equations to position the $k$ to be removed at the top, and write ${\mathbf V}$ in blocks as:
\begin{equation}
{\mathbf V}=
    \begin{bmatrix} 
     \mathbf E  &  \mathbf C' \\
     \mathbf C & \mathbf{V}_{- k} \\
     \end{bmatrix},
 \label{eq:sigmablock}
\end{equation}
where ${\mathbf E}$ is the $k \times k$ block corresponding to the $k$ dropped equations.
%
Let $\mathbf{S_E = E - C' \mathbf{V}_{- k}^{-1}C}$ be Schur-complement of $\mathbf E$. 
Then
${\mathbf V}^{-1}$ can be written as
\begin{equation} 
{\bf V^{-1}}=
\begin{bmatrix}
     {\bf F}  &  {\bf L'} \\
     {\bf L} & {\bf M} \\
     \end{bmatrix}=
    \begin{bmatrix} 
     {\bf S_E^{-1}}  &  {\bf -S_E^{-1} C' \mathbf{V}_{- k}^{-1}} \\
     {\bf -\mathbf{V}_{- k}^{-1} C S_E^{-1}} & {\bf \mathbf{V}_{- k}^{-1} + \mathbf{V}_{- k}^{-1}CS_E^{-1}C' \mathbf{V}_{- k}^{-1}} \\
     \end{bmatrix}
     \label{eq:sigmam1block}
\end{equation}   
\citep{zhang2006schur, haynsworth1968schur}, 
where $\mathbf{F, M, \mbox{ and } L}$ are blocks of the baseline precision matrix 
that are known since we inverted ${\bf V}$ to obtain the current velocity predictions.
(Note that if $\mathbf{V}$ is invertible, then all its leading principal minors are positive and invertible, 
which implies that $\mathbf{S_E}$ and $\mathbf{V}_{- k}$ are also invertible in eq.~\ref{eq:sigmam1block}
is defined.)
Identifying terms in eq.~\ref{eq:sigmam1block} yields:
\begin{align}
\notag \mathbf M &=  \mathbf{V}_{- k}^{-1} + \mathbf{V}_{- k}^{-1}  \mathbf C  \mathbf S_{\mathbf E}^{-1} \mathbf C'\mathbf{V}_{- k}^{-1} \\
\notag &= \mathbf{V}_{- k}^{-1} + \mathbf V_{- k}^{-1}\mathbf C \mathbf S_{\mathbf E}^{-1} \, (\mathbf S_{\mathbf E} \mathbf S_{\mathbf E}^{-1}) \, \mathbf C' \mathbf{V}_{- k}^{-1}\\ 
\notag &= \mathbf{V}_{- k}^{-1} + \mathbf{ L S_E L'}\\ 
\mbox{so that } \mathbf{V}_{- k}^{-1} &= \mathbf{ M - L F^{-1} L'} 
\label{eq:dmk},
\end{align}  
where $\mathbf{F = S_E^{-1}}$ is a $k \times k$ matrix that is easily invertible since $k$ is typically small, and 
that reduces to a scalar when $k=1$.

When we add $k$ equations to the model, we can similarly write:
\begin{equation}
\mathbf V_{+ k} =
    \begin{bmatrix} 
     \mathbf E  &  \mathbf C' \\
     \mathbf C & \mathbf V \\
     \end{bmatrix},
     \label{eq:sigmablock}
\end{equation}
where ${\mathbf E}$ is the $k \times k$ block corresponding to the $k$ added equations.
We apply Shur theorem to obtain the required matrix inverse:
\begin{equation} 
{ \mathbf V_{+k}^{-1}}=
    \begin{bmatrix} 
     {\bf S_E^{-1}}  &  {\bf -S_E^{-1} C' \mathbf{V}^{-1}} \\
     {\bf -\mathbf{V}^{-1} C S_E^{-1}} & {\bf \mathbf{V}^{-1} + \mathbf{V}^{-1}CS_E^{-1}C' \mathbf{V}^{-1}} \\
     \end{bmatrix}
     \label{eq:sigmap1block}
\end{equation}   
where ${\mathbf C}$ and $\mathbf{V}^{-1}$ are known, so that the $k \times k$ Schur-complement $\mathbf{S_E = E - C' \mathbf{V}^{-1}C}$ and the matrix blocks can be evaluated efficiently since all components are known;
we pay only matrix multiplication costs. 
%
%
%

The same short-cut formula applies if we use an $AR(2)$ state equation model:
$
\mathbf{k}_t = A_1 \mathbf{k}_{t-1} + A_2 \mathbf{k}_{t-2} + \eta_t,
$
in place of the $AR(1)$ model in eq.~\ref{StateEq}.
Indeed we let $\mathbf{k^*}_t = (\mathbf{k}_t , \mathbf{k}_{t-1})$, rewrite the $AR(2)$ state equation as an $AR(1)$ process:
$$
\mathbf{k^*}_t
=
\begin{bmatrix} 
    \mathbf A_1  &  \mathbf A_2 \\
    \mathbf I & \mathbf 0 \\
\end{bmatrix}
\mathbf{k^*}_{t-1}
+ 
\begin{bmatrix} 
    \mathbf 1 \\
    \mathbf 0\\
\end{bmatrix}
\eta_t 
$$
and the observation model in eq.~\ref{ObsModFull} as a function of $\mathbf{k^*}$:
$
\mathbf c_{t} = \beta + [\mathbf B, 0]  \mathbf{k^*}_t  + \zeta_t
$
and apply the short-cut formulae to $\mathbf{k^*}_t$. This can further 
be generalized for $AR$ state equation of any order.

\section{Results}
\label{sec:results}

Using the reaching task experiment described below, we illustrate that including waveform features indiscriminately
in the joint decoding model can contribute more noise and bias than useful information about the kinematics, and that
searching for low risk models through a large model space can boost decoding performance substantially. 
We focus primarily on Bayesian predictions
because conclusions for OLE predictions are confounded with another effect: low risk OLE models 
tend to include many observation equations that serve to smooth kinematic predictions \citep{Todorova2017}, 
so we cannot detangle whether adding waveform 
features in an OLE decoder reduces the risk because they contain additional kinematic information 
or because they smooth the predictions.

\begin{figure}
\includegraphics[scale=0.5]{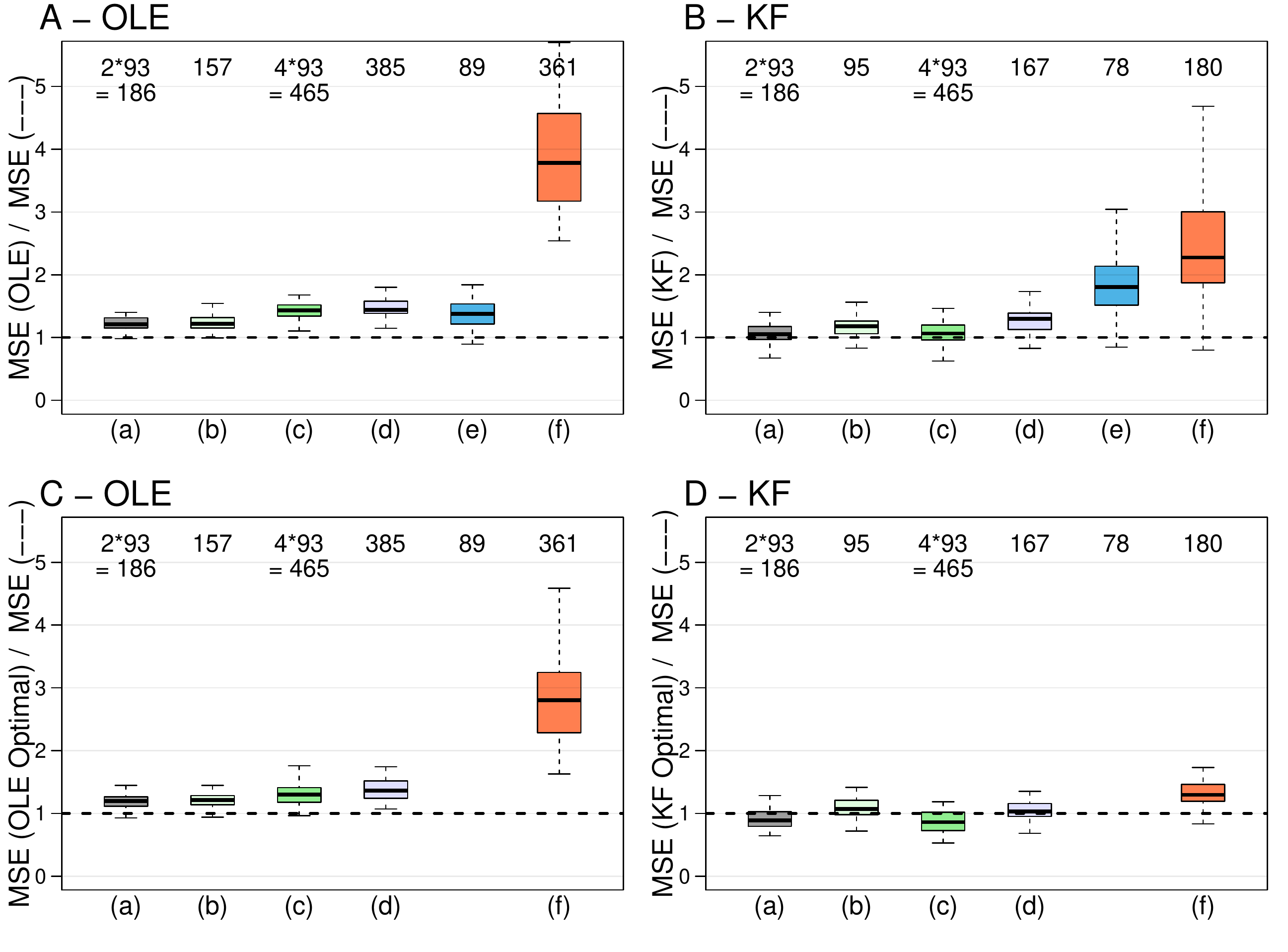}
\caption{\sl Relative efficiencies of various OLE (in A and C) and Bayesian (in B and D) decoding models, relative to two benchmark models: the basic (in A and B) and optimal (in C and D) spike count based OLE and Bayesian models. Each boxplot 
summurizes the relative efficiencies of the 52 test trials.
A relative efficiency of 1.23, say, means that the competing model is 23\% more efficient to decode a trajectory 
than the benchmark model.
The numbers at the top of each panels is the number of observation equations in the various models.
The competing models are composed of (Aa): Eqs.~\ref{ObsMod} and~\ref{ObsModW} with $\tau=8, m=1, l=1$; (Ab): same as (Aa) but pruned; (Ac): Eqs.~\ref{ObsMod} and~\ref{ObsModW} with $\tau=8, m=1, l=1, \ldots, 4$; (Ad): same as (Ac) but pruned;
(Babcd): same as (Aabcd) together with Eq.~\ref{StateEq};
(Cabcd) and (Dabcd): same as (Aabcd) and(Babcd) but observation equations have electrode specific lags and transformations.
(Ae) and (Be): Optimal spike count OLE and Bayesian models with electrode specific lags and transformations.
(Af), (Bf), (Cf), and (Df): optimal joint OLE and Bayesian models.}
\label{fig:stepwise_mle_3D-1}
\end{figure}

\begin{figure}[h]
\includegraphics[scale=0.4]{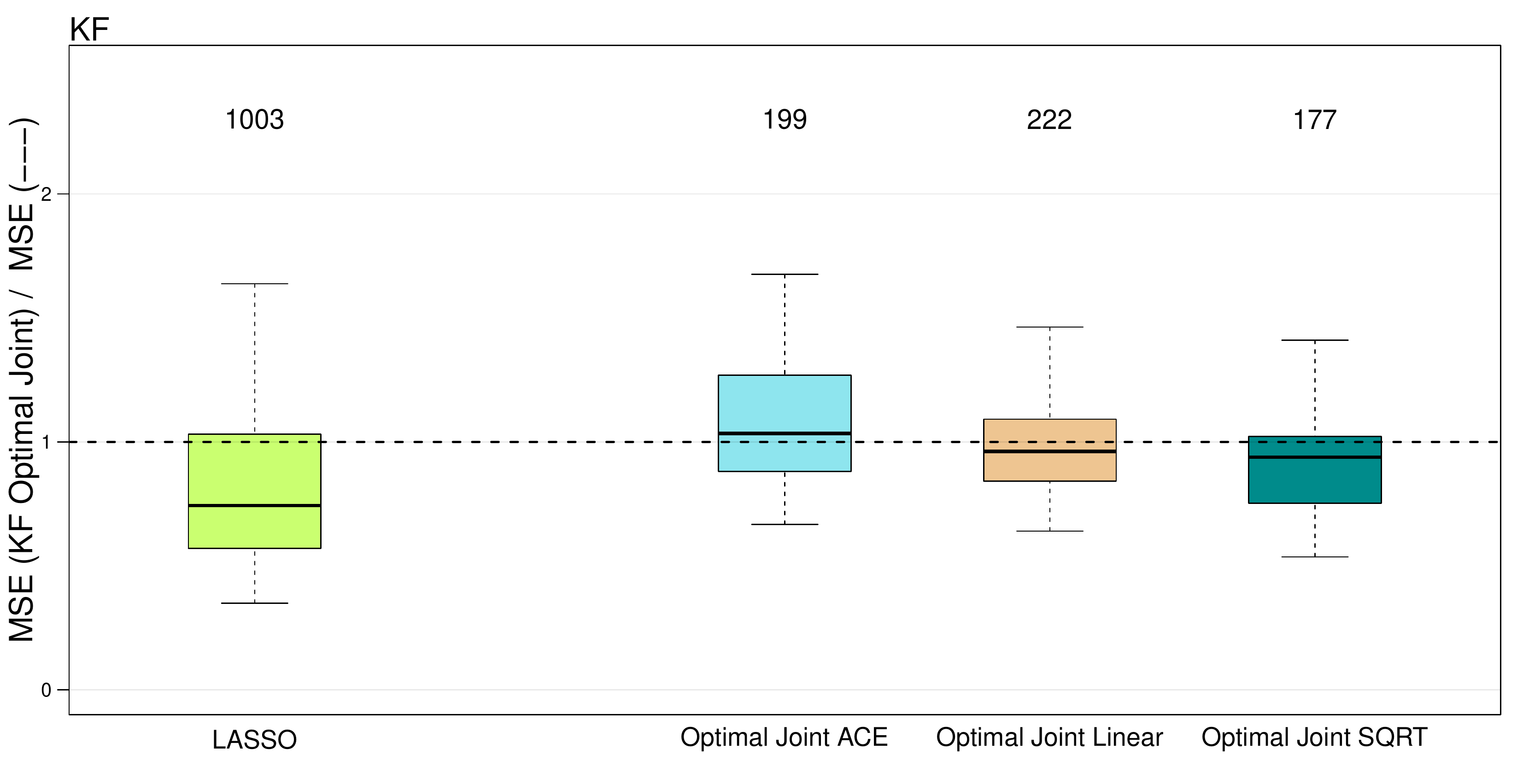}
\caption{\sl 
Efficiencies of LASSO risk minimized forward filter model and 
stepwise Bayesian joint models optimized in smaller model spaces,
relative to the optimal joint Bayesian model. The numbers atop each boxplot are the number of neural covariates 
used in the corresponding models.
The first model is a forward filter that predicts the kinematics as linear combinations of all 13 lagged
raw and transformed electrode spike counts and waveform feature moments, with transformations 
the square root and splines; the model is LASSO optimized to minimize the cross-validated 
prediction risk and drops all but 1003 neural covariates.
The other three models are optimal joint Bayesian models obtained like the benchmark model, but 
optimized within smaller model spaces composed of observation equations that are only untransformed, or
only square root transformed, or only ACE transformed. These three model spaces contains only a third of
the observation equations included in the model space we considered in the main text.}
\label{fig:lasso}
\end{figure}

The neural data were recorded in the primary motor cortex (M1) of a rhesus 
macaque on the $93$ active channels of a $96$-electrode Utah array.
The data includes $10$ sessions, each consisting of $26$ reaches from the center of a virtual 3D sphere 
to 26 targets evenly arranged on a sphere.
Details are in \citep{rasmussen2017dynamic}.
We analyze the portion of the reaches between movement onset and target acquisition,
which amounts to about 8 minutes of data. 
We decode from threshold crossings without spike-sorting, using 16 ms spike count bins.
We initialize the decoders at the observed initial velocity for each trial.
We use eight sessions to train the decoding models and the two remaining 
sessions (${2\times 26}=52$ trials) as the testing set to report model performances in Figs.~\ref{fig:stepwise_mle_3D-1} and~\ref{fig:lasso}.
We cross-validate models, using six of the eight training sessions to estimate the observation equations
and the other two to estimate the risk.
We measure the relative efficiency of two decoders to reconstruct a test trial by the ratio of their respective mean squared errors (MSE) and summarize the distribution of MSE ratios across the $52$ test 
trials using box plots.

%

We first consider the {\bf basic spike count} OLE and Bayesian models composed of the equations in eq. \ref{ObsMod} for the $n=93$ electrodes, lagged by $\tau = 8$ bins (128 ms) compared to arm movement, which
is the best uniform lag in $[0,12]$ that maximizes the average 
$R^2$ of the equations in eq.\ref{ObsMod}.
It is a common model choice that we use as benchmark to compare to other models in Figs.~\ref{fig:stepwise_mle_3D-1}A and~\ref{fig:stepwise_mle_3D-1}B.

\paragraph{Effect of selecting waveform feature equations}

We next illustrate the effect of adding to a decoding model, waveform features with and without 
model selection.
Fig.~\ref{fig:stepwise_mle_3D-1}A(a) and~\ref{fig:stepwise_mle_3D-1}B(a) show the relative efficiencies 
of the 52 test trials decoded using the benchmark basic spike count OLE and Bayesian models together with the observation 
equations for the waveform amplitudes (eq.~\ref{ObsMod2} with $l=1$, $m=1$, and $\tau = 8$).
Relative efficiencies greater than one mean that the benchmark model is less efficient.
Fig.~\ref{fig:stepwise_mle_3D-1}A(b) and~\ref{fig:stepwise_mle_3D-1}B(b) show the relative efficiencies 
of the same models, but with waveform equations pruned to minimize the prediction risk. Note that
different equations may be pruned from the OLE and Bayesian models because the risk of a Bayesian model
depends on the kinematic 
model in eq.~\ref{StateEq}.
Adding all the linear amplitude equations improved spike count based Bayesian predictions
by 5\% in median across the test trials (the median MSE ratio of 1.05 in Fig.~\ref{fig:stepwise_mle_3D-1}B(a)
means that the competing model is 5\% more efficient in median
than the benchmark model). This gain increased to 18\% after pruning 186-95= 91 equations,
which illustrates that selecting which observation equations enter the model is beneficial.
Fig.~\ref{fig:stepwise_mle_3D-1}B(c,d) 
show the corresponding results when we supplement the benchmark basic spike count Bayesian model with
the linear observation equations of the first moment of the four waveform features in Fig.\ref{fig:wf_feat}
(eq.~\ref{ObsMod2} with $l=1, \ldots, 4$, $m=1$, and $\tau = 8$).
Using all four features provides an efficiency gain of just 6\% compared to 5\% using only waveform amplitudes.
However, after pruning 465-167= 298 equations, the relative efficiency increased to 30\%, 
illustrating that the four features together contain more kinematic information than the aplitudes, 
but also more bias and variance, and that pruning reduces the excess bias and variance.

For OLE predictions, adding the four features' first moments to the basic spike count OLE model increased the 
relative efficiency by 63\%, compared to 21\% when adding only the amplitude, a much 
larger discrepancy than with Bayesian predictions.
Pruning these models removed only 80 and 29 equations, respectively, without gaining the sort of 
efficiencies we observed with Bayesian predictions (from 63 to 66\% and 21 to 22\%). 
Both phenomena occur primarily because the additional observation equations improve the prediction risk 
by smoothing the kinematics.
\cite{Kloosterman2013} used an OLE setup to decode a rat position from tetrode recordings
and observed that including two instead of one of the four tetrode amplitudes
increased decoding performance by 26\%, and by a further 14\% when the four tetrode amplitudes 
were included in the decoder. Tetrodes typically record many neurons so it is possible the 
improvement is due to including more 
information, but we cannot tell for certain.

Figs.~\ref{fig:stepwise_mle_3D-1}C(a,b,c,d) and~\ref{fig:stepwise_mle_3D-1}D(a,b,c,d) show the corresponding results when we use the optimal spike count models as benchmarks, which have
electrode specific lags and transformations rather than the linear tuning curves lagged by the same $\tau$ in eq.~\ref{ObsMod}. We describe how to obtain them below.
We supplemented the new benchmarks with the first moments of one or four waveform features,
using the corresponding electrode specific lags and transformations, and evaluated their performances 
before and after they were pruned.
We draw qualitatively stronger conclusions yet:
adding waveform information without model selection to the optimal spike count Bayesian model actually degrades their performance. Pruning the models pushes the 
median relative efficiencies back above one.
Just as above, OLE models tend to improve with more observation equations, simply because they help smooth predictions.

\paragraph{Benefits of searching widely for low risk models}

The first such models are the {\bf optimal spike count} OLE and Bayesian models that serve as benchmarks in figs.~\ref{fig:stepwise_mle_3D-1}C and~\ref{fig:stepwise_mle_3D-1}D.
They were obtained as follows: given OLE or Bayesian paradigm,
we initialized the search at the basic spike count model, and for each electrode in turn,
we determined which of the 3x13 lagged and transformed spike count equations among eq.~\ref{ObsMod2} 
with $g(s)=s, \sqrt{s}$ and $ACE(s)$
reduced the prediction risk the most, and replaced the current with the optimal equation.
We repeated this process until the model stabilized, and then pruned it.
We stepwise searched for the best equation for each 
electrode in turn because most electrodes contribute kinematic information, and performing a single stepwise 
search across all equations at once does not explore the model space as fully and runs the risk of excluding 
important electrodes if the search is stuck in a local minimum risk part of the space.
The choice of initial model and the order for updating electrodes had no effect on 
the quality of the final models.

The resulting optimal spike count Bayesian model 
is 81\% more efficient in median than its basic counterpart (Fig.~\ref{fig:stepwise_mle_3D-1}B(e)), which further validates the findings in \citep{Wu2006, Todorova2017} that 
using electrode specific lags and transformations can improve decoding performance.
The corresponding efficiency gain in the OLE framework is 37\% (Fig.~\ref{fig:stepwise_mle_3D-1}A(e)), which is much less because
the number of equations in the model (89) is too few to smooth the predictions.

The {\bf optimal joint} OLE and Bayesian models are the outcomes of our most extensive 
model search: 
we searched for minimum risk electrode specific lags and transformations models for each of the eight waveform 
moments
using the same procedure that yielded the optimal spike count models, 
combined each of these eight models in turn
with the optimal spike count model, pruned each of these eight combined models over waveform equations only, combined all the remaining unpruned waveform equations with the opimal spike count model, and pruned that 
model. 
We searched separately for optimal spike count and optimal waveform feature models because 
they provide 
different kinematic information: the former provides information combined across the neurons they
record, and the latter retrieve the additional information that each of these neurons provide, 
without spike sorting them; we further treat the different waveform features separately 
because different features may best separate different neurons \citep{Ventura2015}.

The optimal joint Bayesian model (Figs.~\ref{fig:stepwise_mle_3D-1}B(f)) is 128\% more
efficient in median than its basic spike count counterpart, which shows the kind of improvement one can expect compared to a
commonly used model, and Fig.~\ref{fig:stepwise_mle_3D-1}D(f) shows that it is 30\% more efficient than its optimal spike count counterpart, which illustrates the amount of additional information one an extract from the 
waveforms, once optimal spike count information is included in the model.
It is also comparably efficient to the 
{\bf optimal sorted spike count and hash} Bayesian models (not shown),
which was obtained in the same manner as the optimal (electrode) 
spike count model but based on the spike trains of 
the sorted neurons together with the hash spike trains from each electrodes..
This confirms the previously reported result that decoding from the joint model is at least as efficient as
decoding from sorted neurons 
and hash, while avoiding the problematic spike sorting step.
The optimal joint OLE model in Figs.~\ref{fig:stepwise_mle_3D-1}A(f) contains 361 equations versus the 180 
equations of the optimal joint Bayesian model. It is much larger because many equations serve to smooth 
the kinematic trajectories. Its efficiency relative to the spike count OLE models is impressive, 
but confound the effect of additional kinematic information with smooth kinematic trajectories.

So far we have compared models within Bayesian and OLE decoding paradigms.
Comparing across paradigms, the Bayesian models are superior.
The efficiency of the optimal spike count Bayesian model, with 78 equations, relative to the optimal spike count 
OLE model, with 89 equations, across the 52 test trials ranges from 120\% to 1490\%, with quartiles 423, 533, and 752\% (not shown);
Bayesian predictions are far superior because the state equation in eq.~\ref{StateEq} provide substantial information 
when the observation model is small.
The corresponding five number summaries to compare optimal joint Bayesian and OLE models (180 and 361 equations, respectively) are minimum 26\%, maximum 538\%, and quartiles 177, 208, and 281\%.

\paragraph{Parametric versus nonparametric models}

The model space we explored includes transformed response 
linear observation equations for all spike counts and electrode feature moments, with response transformations 
the identity
(eqs.~\ref{ObsMod} and~\ref{ObsModW}), square
root, and ACE response transformations (eq.~\ref{ObsMod2} with $g(\cdot)=\sqrt{\cdot}$ and $g=ACE(\cdot)$).
%
We also obtained optimal joint models in smaller model spaces composed of
identity equations only, square root equations only, and ACE equations only. These spacesare three times smaller
so they are faster to explore.
Fig.~\ref{fig:lasso} in Appendix shows the efficiencies of the three resulting Bayesian models relative
to the the global optimal joint Bayesian model: 
the efficiencies of these four models are within 4\% of one another, 
the square root response transformations only model being the least efficient and the 
the ACE response transformations only model the most.
For this dataset, and perhaps for most, we could afford to work only with ACE transformed linear observation 
equations.

\paragraph{Forward filter decoding models}

Model selection is straightforward for forward filters that predict kinematics as functions of neural covariates, using 
off the shelf tools like stepwise regression, LASSO, or ridge regression.
\cite{Todorova2017} illustrate that risk minimized forward 
filters can improve decoding accuracy very substantially, but argue that a risk optimized Bayesian decoder may be more efficient yet,
although they could not investigate the claim fully because model searches
were computationally too expensive.
Here, we were able to conduct a more extensive search with our short cut formula in eq.~\ref{eq:dmk}.
Fig.~\ref{fig:lasso} in Appendix shows that our optimal joint Bayesian 
model is 35\% more efficient in median across the testing set than the corresponding LASSO risk optimized forward filter that predicts 
kinematics as linear combinations of untransformed, square root transformed, and splines transformed electrode spike counts
and waveform feature moments at all 13 lags. (We included 351 spike counts and waveform moments per electrode in the model;
LASSO eliminates all but 1003 of these covariates.)

\paragraph{Computational efficiency of the short cut formula}

\begin{figure}[t!]
\centerline{ \includegraphics[scale=0.4]{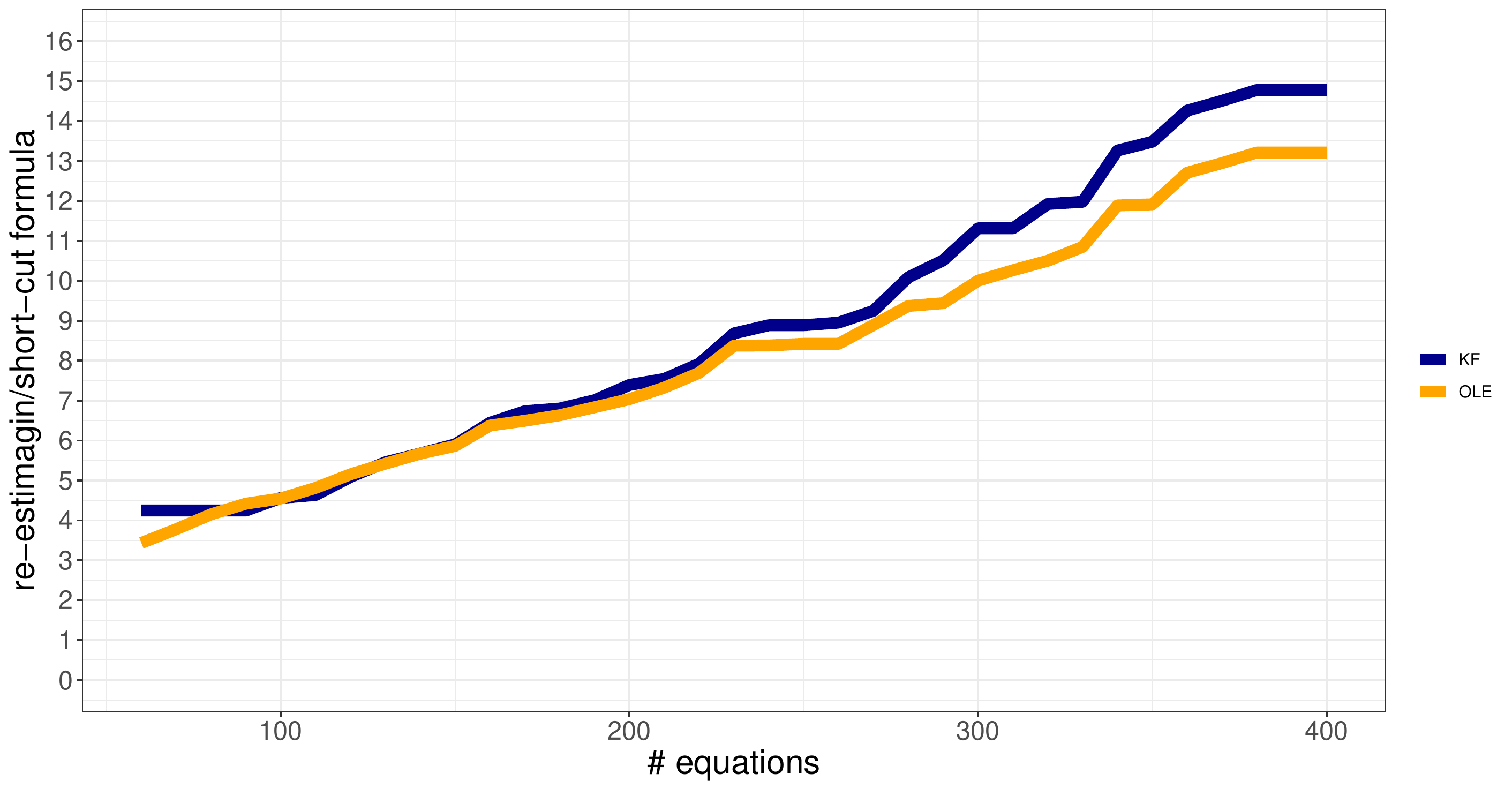} }
\caption{\sl Relative computational efficiency of choosing which equation to leave out in a model of size $N$, using the short cut formula  in eq.~\ref{eq:dmk} for inverting matrices comparing to inverting them directly. The short cut formula is faster by a factor proprtional to $N$. }
\label{fig:shortcut_efficiencty}
\end{figure}

Given a number $N$ of equations in a decoding model, we created the $N$ leave-one-equation-out submodels , and for each, obtained OLE and Bayesian predictions for the 52 test trials, 
either using the sort cut formula in eq.~\ref{eq:dmk} to obtain the inverse of $\mathbf V_{-k}$, with $k=1$, given that $\mathbf V^{-1}$
in eq.~\ref{eq:sigmam1block} is known, or inverting $\mathbf V_{-k}$ directly.
Fig.~\ref{fig:shortcut_efficiencty} shows the ratio of the user CPU times of the second over the first inversion method
as a function of $N$, using
the programming language R, CPU Quad-core 1.8 GHz Intel Core i5, and RAM 16 GB 2133 MHz DDR4.
The relative computational demand for both OLE and Bayesian predictions are linear in $N$, as we expected
since the short cut formula has complexity $O(N^2)$ and inverting a matrix has complexity $O(N^3)$.
We see that when $N= 200$, say, the short cut formula is 7 times faster.
When pruning the first equation out of the models with $N=465$ observation equations in 
Figs.~\ref{fig:stepwise_mle_3D-1}A(d), \ref{fig:stepwise_mle_3D-1}B(d), \ref{fig:stepwise_mle_3D-1}C(d), and \ref{fig:stepwise_mle_3D-1}D(d), the short cut formula is well over 15 times faster, by interpolating the curve in 
Fig.~\ref{fig:shortcut_efficiencty}. We expect that the short cut formula might be more efficient yet 
(have a steeper slope)
by using a lower level language such as C++ that has better working space management.

\section{Discussion}

A recent paradigm for decoding kinematics or other stimuli from neuron ensembles 
relies on joint models for electrode spike counts and waveform features,
which, in principle, is more efficient than decoding from unsorted spike trains alone or from sorted spike trains. 

In this paper, we showed that including 
waveform features indiscriminately in the joint decoding model can contribute more noise and bias than useful 
information about the kinematics, and thus degrade decoding performance, and that
selecting which waveform features should enter the model
can boost decoding performance substantially.
For the data analyzed here, a stepwise search within a model space composed of
electrode spike count and waveform moment observation equations
yielded a low risk Bayesian joint model that is 
30\% more efficient than the corresponding risk minimized Bayesian model based only on electrode spike 
count information. 
The joint model is also comparably efficient to decoding from a risk minimized model based only on 
sorted neuron spike trains and hash, confirming previous results that one can do away with the 
problematic spike sorting step in decoding applications.
Finally, our optimal Bayesian spike and waveform joint decoder is also 35\% more efficient 
than a forward filter that uses the same information and that is fitted to the data using 
LASSO to minimize the prediction risk.

We were able to search through a large model space thanks to the proposed 
short cut formula, which accelerates matrix inversions in stepwise searches
for models based on Gaussian linear observation equations, in the context of 
decoding neural ensembles or in the many other situations where state-space models
are used.

\section*{Acknowledgements}
VV is supported by NIH grant R01 MH064537.  We are grateful to Steve Chase and Lindsay Bahureksa
for data collection and assistance with data processing.

\section*{Appendix}

\clearpage


\bibliographystyle{apalike}  
\bibliography{SpikeLit}

\end{document}